\documentclass[aps,preprint,nofootinbib,superscriptaddress,prc]{revtex4}
\usepackage{amssymb}

\usepackage{epsfig}
\usepackage[bookmarksnumbered,bookmarksopen,colorlinks,citecolor=blue,linkcolor=blue]{hyperref}
\begin{document}


\title{ Nuclear symmetry energy from the Fermi-energy difference in nuclei}

\author{Ning Wang}
\email{wangning@gxnu.edu.cn}
\affiliation{Department of
Physics,Guangxi Normal University, Guilin, 541004, P. R. China}

\author{Li Ou}
\affiliation{Department of Physics,Guangxi Normal University,
Guilin, 541004, P. R. China}

\author{Min Liu}
\affiliation{Department of Physics,Guangxi Normal University,
Guilin, 541004, P. R. China}

\fontsize{12pt}{12pt}\selectfont

\begin{abstract}
The neutron-proton Fermi energy difference and the correlation to
nucleon separation energies for some magic nuclei are investigated
with the Skyrme energy density functionals and nuclear masses, with
which the nuclear symmetry energy at sub-saturation densities is
constrained from 54 Skyrme parameter sets. The extracted nuclear
symmetry energy at sub-saturation density of 0.11 fm$^{-3}$ is 26.2
$\pm$ 1.0 MeV with 1.5$\sigma$ uncertainty. By further combining the
neutron-skin thickness of $^{208}$Pb, ten Skyrme forces with slope
parameter of $28 \leqslant L \leqslant 65$ are selected for the
description of the symmetry energy around saturation densities.
\end{abstract}


\maketitle

\begin{center}
\textbf{I. INTRODUCTION}
\end{center}

The nuclear symmetry energy, in particular its density dependence,
has received considerable attention in recent years
\cite{Zhang08,Tsang09,Li08,Li04,Chen05,Shet07,Botvina02,Cent09,Stein,Stein05,Dong11,Dong12}.
As one of the key properties of nuclear matter, the nuclear symmetry
energy probes the isospin part of nuclear force and intimately
relates to the structure character of drip line nuclei and
super-heavy nuclei, the dynamical process of nuclear reactions and
the behavior of neutron stars. To explore the density dependence of
the nuclear symmetry energy from sub-saturation to super-saturation
densities, various models and experimental observables have been
proposed. On one hand, the constraints on the symmetry energy are
investigated from heavy-ion collisions \cite{Xiao,Rus11,Tsang09}.
Some experimental data for the isospin sensitive observables are
well reproduced by using certain forms of density-dependent symmetry
energy in microscopic dynamics calculations, such as in the improved
quantum molecular dynamics  \cite{Tsang09,Zhang08} and the isospin
Boltzmann-Uehling-Uhlenbeck \cite{Li04,Chen05} calculations. In
these calculations, the temperature effect and the influence of the
isospin-independent terms of nuclear force are self-consistently
involved. It is still difficult to clearly obtain the information of
nuclear symmetry energy at zero-temperature by removing the
influence of the isospin-independent terms.

On the other hand, the symmetry energy is also constrained from the
properties of finite nuclei, such as the binding energies
\cite{Dani09,Wang10,Liu10,Chen11,FRDM12,Diep09}, the neutron skin
thickness \cite{Cent09,Warda09,Roca,Tam11,Kras12} and the pygmy
dipole resonance (PDR) \cite{Klim07,Carb10}. By analyzing the more
than 2000 measured masses of nuclei with the help of the liquid drop
formula, one can obtain the mass dependence of the symmetry energy
coefficients of finite nuclei \cite{Dani09,Wang10,Diep09} and the
nuclear symmetry energy at sub-saturation densities
\cite{Liu10,Chen11} based on the relation between the symmetry
energy coefficients of finite nuclei and the symmetry energy of
infinite nuclear matter \cite{Cent09}. The obtained symmetry energy
at the saturation density and its slope parameter \cite{Liu10} are
generally close to the results from PDR \cite{Carb10} and heavy-ion
collisions \cite{Zhang08}. In addition, the symmetry energy is also
constrained from neutron star observations incorporates the
microphysics of both the stellar crust and core
\cite{Stein,Wen12,Latt12}. A relatively smaller slope parameter of
the symmetry energy at the saturation density $43<L<52$ MeV is
obtained from the available neutron star mass and radius
measurements \cite{Stein}. It is known that the average density of a
finite nucleus is smaller than the saturation density due to the
surface diffuseness. The obtained information mainly describes the
nuclear symmetry energy at the sub-saturation densities rather than
the super-saturation densities. The symmetry energies at the
saturation and super-saturation densities based on the extrapolation
are still very uncertain and more isospin-sensitive obervations
should be proposed and investigated.

The density dependence of the symmetry energy is also extensively
studied with the Skryme energy density functionals
\cite{Chen12,HFB17,Dut12}. The Hartree-Fock-Bogoliubov approach with
the Skyrme force BSk17 \cite{HFB17} can reproduce the 2149 measured
masses with an rms deviation of 0.581 MeV, which is comparable to
the accuracy of the new finite range droplet model (FRDM2012)
\cite{FRDM12}. In the FRDM2012, it is found that a slope parameter
of the symmetry energy at normal density $L \approx 70 $ MeV can
give better results with an rms deviation of 0.570 MeV. However, one
should note that the corresponding slope parameter  is only $L=36$
MeV from the BSk17 parameter set which is much smaller than the
result of FRDM2012. It is therefore necessary to further investigate
the behavior of the symmetry energy at sub-saturation densities from
the structures of finite nuclei.

In this work, we study the nuclear symmetry energy at sub-saturation
densities from the Fermi energies of nuclei based on various
parametrizations of the Skyrme forces. The Skyrme interaction,
originally constructed for finite nuclei and nuclear matter at
saturation density, is a low momentum expansion of the effective
two-body NN interaction in momentum space. Although all Skyrme
forces are usually fitted to reproduce well the saturation energy
and density of symmetric nuclear matter, they differ significantly
in other characteristics of symmetric and pure neutron matter, in
particular their density dependence \cite{Dut12}. The Fermi energies
of neutrons and protons for some doubly-magic nuclei can be measured
with a high precision. The neutron-proton Fermi energy difference
and the correlation to nucleon separation energies of magic nuclei
are directly related to the symmetry energy of nuclei, with which
the isospin-dependent terms of the Skyrme forces and thus the
density dependence of the symmetry energy could be constrained. The
paper is organized as follows: In Sec. II, the correlation between
Fermi energies and separation energies of nucleons are introduced.
In Sec. III, the nuclear symmetry energy at sub-saturation and
saturation densities is extracted. Finally, a summary is given in
Sec. IV.

\begin{center}
\textbf{II. CORRELATION BETWEEN FERMI ENERGIES AND SEPARATION
ENERGIES OF NUCLEONS}
\end{center}

 Based on the liquid drop mass formula, the binding energy of a nucleus which
 is taken as positive value is expressed as
\begin{eqnarray}
   BE(A,Z)=a_{v} A - a_{s} A^{2/3}- a_{c} \frac{Z^2}{A^{1/3}} - a_{\rm
   sym}I^2 A
\end{eqnarray}
by neglecting nuclear microscopic corrections. $I=(N-Z)/A$ denotes
the isospin asymmetry. The difference between the proton separation
energy $[BE(A,Z)-BE(A-1,Z-1)]$ and the neutron separation energy
$[BE(A,Z)-BE(A-1,Z)]$ of a nucleus is written as,
\begin{eqnarray}
   \Delta S=S_p-S_n  \simeq -2 a_{c} \frac{Z}{A^{1/3}} + 4 a_{\rm
   sym} I.
\end{eqnarray}
Because the Coulomb energy coefficient $a_c=\frac{3}{5}\frac{ e^2}{
r_0}\simeq 0.71$ MeV is usually well determined from the masses of
mirror nuclei \cite{Wang10a,Liu11}, the value of $\Delta S$ is
directly related to the symmetry energy coefficients of finite
nuclei.

On the other hand, the single-particle energies (SPE) of a nucleus
can be uniquely determined by solving the Schr\"odinger equations or
Hartree-Fock equations based on the single particle potential, under
the mean-field approximation. In the Hartree-Fock theory for a
closed-shell nucleus $(A,Z)$ the single-particle energies for states
below the Fermi surface are given by \cite{Brown,Schwierz}
\begin{eqnarray}
   \varepsilon_p=BE^*(A-1,Z-1)-BE(A,Z)
\end{eqnarray}
and
\begin{eqnarray}
   \varepsilon_n=BE^*(A-1,Z)-BE(A,Z).
\end{eqnarray}
$\varepsilon$ will be negative for bound states. $(BE^* = BE-E_x)$
is the ground state binding energy minus the excitation energy of
the excited states associated with the single-particle states. The
difference between the Fermi energy of neutrons and that of protons
\begin{eqnarray}
   \Delta \varepsilon=\varepsilon^F_n-\varepsilon^F_p
\end{eqnarray}
is closely related to the proton and neutron separation energies of
the nucleus.  Here, the Fermi energy $\varepsilon^F$ is defined as
the energy of the highest occupied quantum state in a system of
fermions at absolute zero temperature.

\begin{figure} 
     \centering
        \includegraphics[width=12cm]{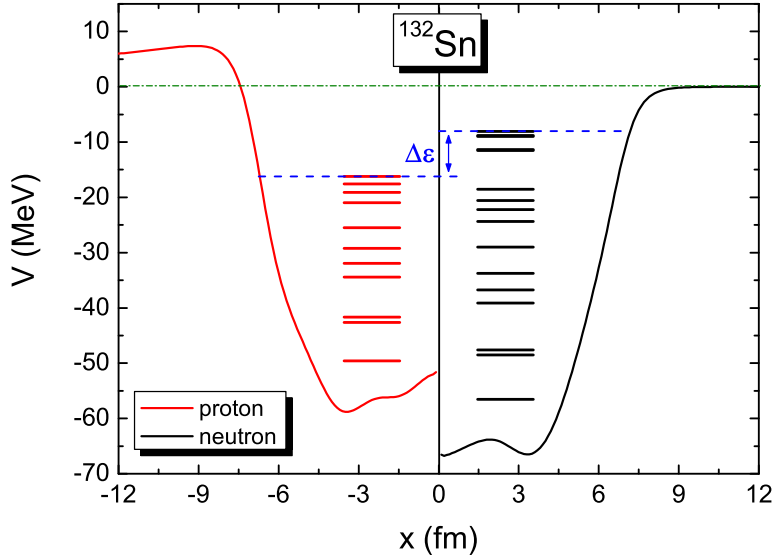}
        \caption{(Color online) Single particle potentials and the single particle energies (SPE) of bound states
        for $^{132}$Sn with the Skyrme Hartree-Fock calculation by using SLy7 force. The dashed lines denote
        the highest SPE of the occupied states for nucleons in $^{132}$Sn at its ground state. }
        \label{fig1}
    \end{figure}

 \begin{figure} 
     \centering
        \includegraphics[width=13cm]{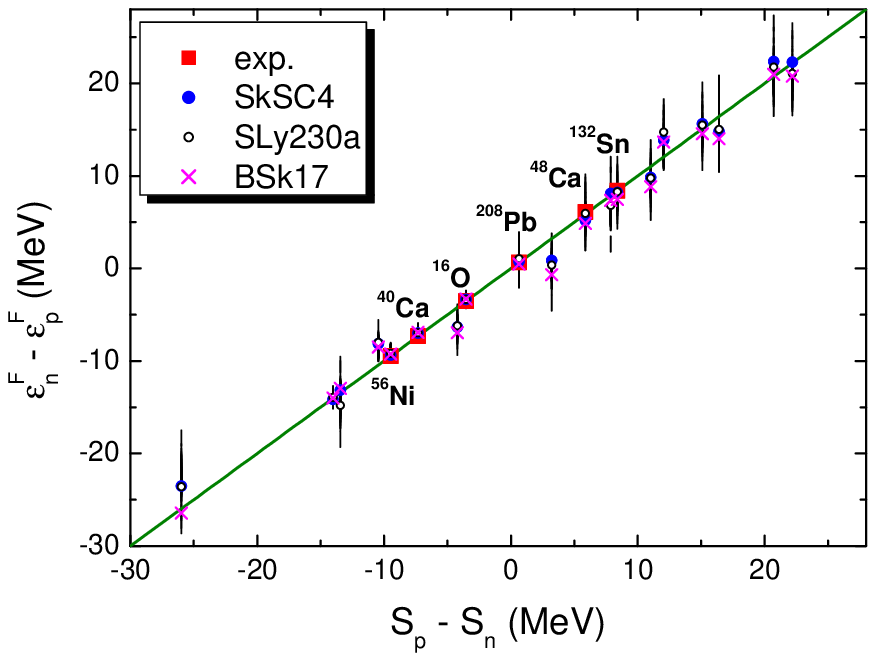}
        \caption{(Color online) Fermi energy difference as a function of separation energy difference. The red squares denote the experimental data for
         six doubly-magic nuclei \cite{Brown,Schwierz}. Others are the calculated results with difference Skyrme forces. }
        \label{fig1}
    \end{figure}

Fig. 1 shows the calculated single particle energies of occupied
states for protons and neutrons of $^{132}$Sn at its ground state by
using the Skyrme Hartree-Fock (SHF) model with the parameter set
SLy7 \cite{Chab98}. For this neutron-rich nucleus, the Fermi energy
of neutrons is higher than that of protons, and the calculated value
of $\Delta \varepsilon$ is 8.1 MeV. The depth of the single particle
potential plays a dominant role for the corresponding Fermi energy
of a given nucleus. The difference $\Delta \varepsilon$ closely
relates to the difference between the depth of nuclear potentials
for neutrons and protons,
\begin{eqnarray}
 V_n-(V_p+V_C) \simeq -\frac{3 }{2 } \frac{e^2}{r_c} \frac{Z}{A^{1/3}} + 2 V_{\rm sym} I.
\end{eqnarray}
Here, $V_C \simeq \frac{3 }{2 } \frac{e^2}{r_c} \frac{Z}{A^{1/3}}$
denotes the Coulomb potential of a nucleus at the central position
with the potential radius $r_c \approx 1.3$ fm. The information on
the symmetry potential $V_{\rm sym}$ from the Fermi energy
difference is of great importance for the study of nuclear symmetry
energy.

It is known that under the mean-field approximation, if the
single-particle motion plays a dominant role for the behavior of
nucleons near the Fermi surface, one expects that the relation
$\Delta S \simeq \Delta \varepsilon$ holds for the closed-shell
nuclei. The experimental values  of $\Delta \varepsilon$ for six
doubly-magic nuclei $^{16}$O, $^{40}$Ca, $^{48}$Ca, $^{56}$Ni,
$^{132}$Sn and $^{208}$Pb are $-3.53$, $-7.31$, $6.10$, $-9.47$,
$8.40$ and $0.64$ MeV, respectively \cite{Brown,Schwierz}. The
corresponding separation energy differences $\Delta S$ for these six
nuclei are $-3.53$, $-7.31$, $5.86$, $-9.48$, $8.39$, $0.64$ MeV,
respectively. One sees that the relation $\Delta S \simeq \Delta
\varepsilon$ does hold very well as expected for these doubly-magic
nuclei. To further test the relation, the values of $\Delta
\varepsilon$ for 19 doubly-magic or semi-magic nuclei ($^{16,22}$O,
$^{22,42}$Si, $^{40,48,60}$Ca, $^{42}$Ti, $^{56,68,78}$Ni,
$^{130}$Cd, $^{100,132,134}$Sn,$^{134}$Te, $^{144}$Sm,
$^{182,208}$Pb) are systematically investigated by using the SHF
model together with 54 commonly used Skyrme forces. The calculated
values are shown in Fig. 2 as a function of $\Delta S$. Here, the
masses of unmeasured nuclei are predicted with the
Weizs\"acker-Skyrme mass formula combining the radial basis function
correction (WS3$^{\rm RBF}$) \cite{Wang11}. The WS3$^{\rm RBF}$
model can reproduce the measured 2149 masses in AME2003 with an rms
deviation of 184 keV, and the predictive power is also remarkable
\cite{Wang12} (the rms deviation with respect to the 154 new masses
of extremely neutron-rich and proton-rich nuclei listed in AME2012
\cite{AME2012} is only 397 keV). The red squares in Fig. 2 which
denotes the experimental data for the six doubly-magic nucleus
mentioned previously are quite regularly located along the green
line $\Delta S = \Delta \varepsilon$. The solid circles, open
circles and crosses denote the results of three Skyrme forces SkSC4
\cite{SKSC4}, SLy230a \cite{SLy}  and BSk17 \cite{HFB17},
respectively. The error bars denote the uncertainty of the model
calculations from 54 different Skyrme forces in which the
corresponding incompressibility coefficient  for symmetry nuclear
matter is $K_\infty=230 \pm 30$ MeV and the saturation density
$\rho_0=0.16\pm0.005$ fm$^{-3}$. One sees that the calculated
results from the traditional 10-parameter Skyrme forces support the
relation $\Delta S \simeq \Delta \varepsilon$, even for the
extremely neutron-rich and proton-rich nuclei. The correlation
between the Fermi energy difference and the separation energy
difference for neutrons and protons could be helpful for
constraining the equation of state for asymmetry nuclear matter.
Here, we would like to emphasize that the Fermi energy difference
can effectively remove the influence of isospin-independent terms in
the nuclear forces.

\begin{center}
\textbf{III. NUCLEAR SYMMETRY ENERGY AT SUB-SATURATION AND
SATURATION DENSITIES}
\end{center}

\begin{figure} 
     \centering
        \includegraphics[width=0.9\textwidth]{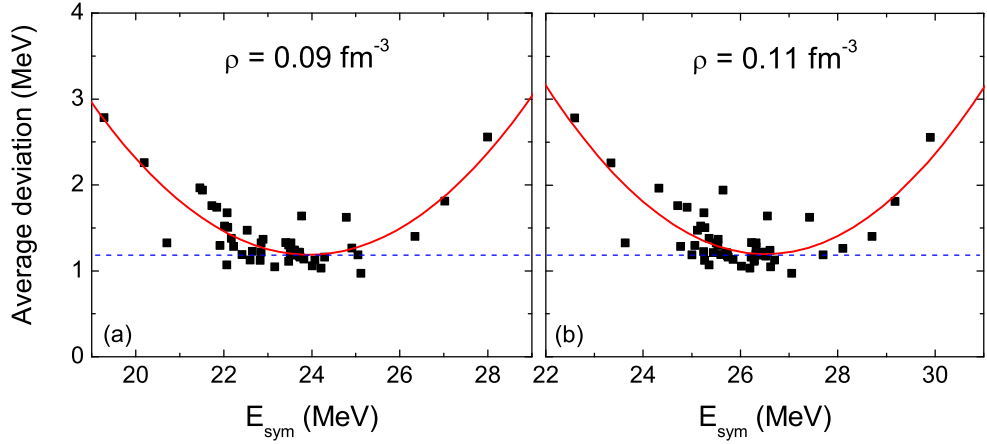}
        \caption{(Color online) Average deviation as a function of nuclear symmetry
energy $E_{\rm sym}$ at the density of (a) $\rho=0.09$ fm$^{-3}$ and
(b) 0.11 fm$^{-3}$, with 54 different Skyrme forces. The solid
curves denote the parabolic fit to the squares. The dashed line show
the position of 1.19 MeV.}
        \label{fig2}
    \end{figure}

Based on the calculated Fermi energy difference $\Delta \varepsilon$
with the 54 different Skyrme forces, the average deviation
\begin{eqnarray}
 \langle \sigma \rangle = \frac{1}{m}\sum^m_{i=1}  {\mid \Delta \varepsilon^{(i)} - \Delta S^{(i)} \mid}
\end{eqnarray}
from the $m=19$ nuclei mentioned previously is calculated. Fig. 3(a)
and (b) show the average deviation as a function of symmetry energy
$E_{\rm sym}$ at the density of $\rho=0.09$ and 0.11 fm$^{-3}$,
respectively. The nuclear symmetry energy in the Skyrme energy
density functional is expressed as
\begin{eqnarray}
E_{\rm sym}(\rho) &=& \frac{1}{2} \left [\frac{\partial^2 (E/A)}{\partial I^2} \right ]_{I=0} \nonumber  \\
&=& \frac{1}{3}\frac{\hbar^2}{2m} \left ( \frac{3\pi^2}{2}
\right )^{2/3} \rho^{2/3}-  \frac{1}{8} t_0 (2 x_0+1)\rho  \nonumber \\
&-&   \frac{1}{24} \left(
 \frac{3 \pi^2}{2} \right )^{2/3} \Theta_{\rm sym}
 \rho^{5/3}-\frac{1}{48} t_3 (2x_3+1) \rho^{\sigma+1}
\end{eqnarray}
with $ \Theta_{\rm sym}=3 t_1 x_1 -t_2(4+5 x_2)$. $t_0$, $t_1$,
$t_2$, $t_3$, $x_0$, $x_1$, $x_2$, $x_3$ and $\sigma$  are the
Skyrme parameters. The squares and solid curves denote the results
of 54 Skyrme forces and the parabolic fit, respectively.  One sees
that the minimal deviations are located at around $E_{\rm
sym}(0.09)= 23.6$ and $E_{\rm sym}(0.11) = 26.2$ MeV, respectively.
We also note that the obtained symmetry energies do not change
appreciately if only the six doubly-magic nuclei are involved in the
calculation of the average deviation $\langle \sigma \rangle$.  It
indicates that the Fermi energy difference is a useful observation
for studying the symmetry energy at sub-saturation densities. The
dashed line shows the minimal value of the parabolic curve which is
1.19 MeV. The minimal average deviation from the 54 Skyrme forces is
about $1.0$ MeV. Considering the systematic error of the WS3$^{\rm
RBF}$ mass model which is about 0.19 MeV \cite{Wang11}, the Skyrme
forces with $\langle \sigma \rangle \le 1.19 $ MeV reasonably well
describe the Fermi energy difference for the 19 nuclei mentioned
previously.

\begin{figure} 
     \centering
        \includegraphics[width=0.7\textwidth]{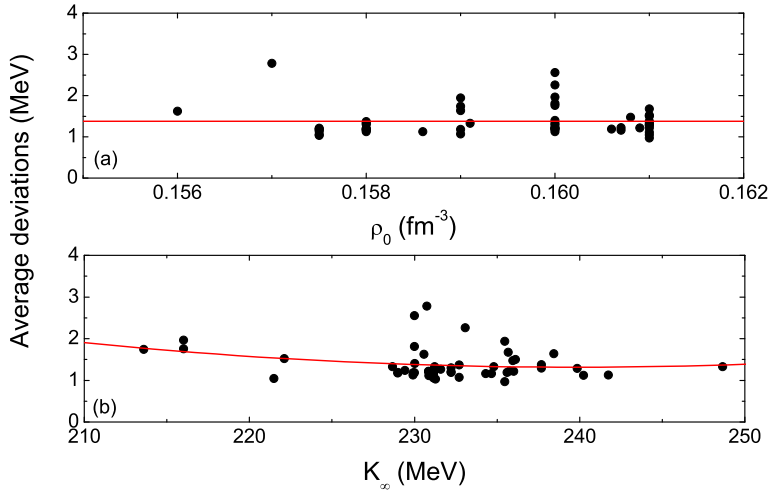}
        \caption{(Color online) Average deviation as a function of (a) saturation density $\rho_0$ and
        (b) incompressibility coefficient $K_\infty$ of symmetric nuclear matter, based on the calculations of the 54 Skyrme forces. }
        \label{fig3}
    \end{figure}

As one of the key properties of nuclear matter, the symmetry energy
is particularly important in modelling nuclear matter and finite
nuclei because it probes the isospin part of the Skyrme interaction.
For a sensitive observation to investigate the nuclear symmetry
energy, the influence of isospin-independent terms in the nuclear
forces should be removed as clean as possible. In Fig. 4, we show
the average deviation as a function of nuclear saturation density
$\rho_0$ and the incompressibility coefficient $K_\infty$ for
symmetry nuclear matter. One cannot evidently obtain the optimal
values of $\rho_0$ and $K_\infty$ according to the average
deviations from the 54 Skyrme forces, since these two quantities are
determined by the isospin-independent parts of the Skyrme
interactions. We also note that the strength of the spin-orbit
interaction $W_0$ in the Skyrme forces does not affect the value of
$\Delta \varepsilon$ generally due to the cancelation between
protons and neutrons, which is helpful to remove the influence of
the shell effect on the symmetry energy. For example, the value of
$\Delta \varepsilon$ for $^{132}$Sn only changes by $0.7\%$ with a
variation of the strength of the spin-orbit interaction by $32\%$
according to the SLy7 calculations.

\begin{figure}
     \centering
        \includegraphics[width=0.9\textwidth]{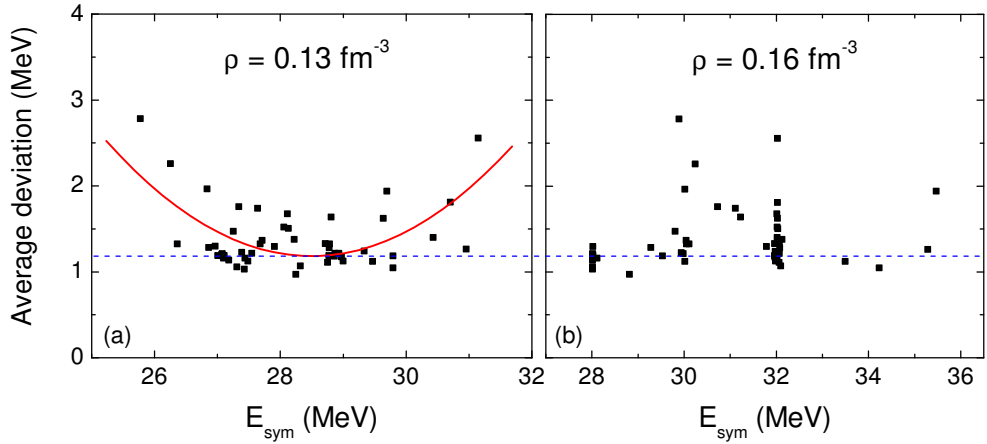}
        \caption{(Color online) The same as Fig.3, but at the density of $\rho=0.13$ and 0.16 fm$^{-3}$, respectively. }
        \label{fig3}
    \end{figure}

From the 54 Skyrme forces, 17 parameter sets with $\langle \sigma
\rangle \le 1.19$ MeV are selected for the description of $E_{\rm
sym} (\rho)$. These selected Skyrme forces can well reproduce the
experimental data for the Fermi energy difference of the six
doubly-magic nuclei. We note that the obtained symmetry energies
from these 17 Skyrme interactions are close to each other at the
densities around $\rho_c=0.11$ fm$^{-3}$, and the value of $E_{\rm
sym} (\rho_c)=26.2 \pm 1.0$ MeV with $1.5\sigma$ uncertainty. It
could be much more useful if the slope parameter of the symmetry
energy at the saturation density $\rho_0$
\begin{eqnarray}
L=3\rho_0 \left (\frac{\partial E_{\rm sym}}{\partial \rho} \right
)_{\rho=\rho_0}
\end{eqnarray} can be also well constrained with this approach.
Unfortunately, we find that the obtained uncertainties of the
symmetry energy at densities lower and higher than $\rho_c$
gradually increase. Fig. 5 shows the average deviation as a function
of nuclear symmetry energy $E_{\rm sym}$ at the density of
$\rho=0.13$ and 0.16 fm$^{-3}$, respectively. The parabolic behavior
of the average deviation becomes unclear with increasing of density
and even disappears at the saturation density. The corresponding
uncertainties of the symmetry energy at $\rho=0.13$ and 0.14
fm$^{-3}$ increase to 1.5 and 2.0 MeV, respectively. It indicates
that the slope parameter $L$ can not be well constrained by using
the neutron-proton Fermi energy difference uniquely, although the
nuclear symmetry energy at sub-saturation density can be well
described.

\begin{figure}
     \centering
        \includegraphics[width=0.7\textwidth]{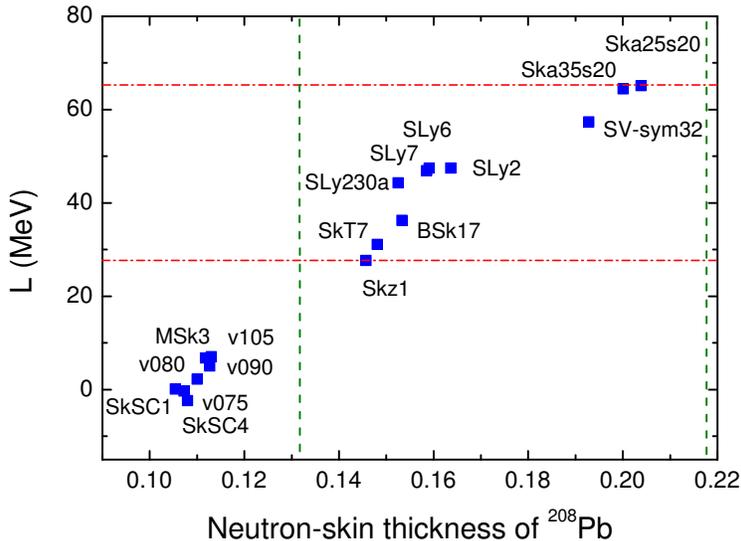}
        \caption{(Color online) Slope parameter of nuclear symmetry energy as a function of neutron-skin thickness of $^{208}$Pb.
        The region between the two dashed vertical lines denotes the measured neutron-skin thickness of $^{208}$Pb,  $0.131 \leqslant
\Delta R_{\rm np} (^{208} {\rm Pb}) \leqslant 0.218 $ fm
\cite{Klim07,Carb10,Tam11,Kras12}.}
        \label{fig3}
    \end{figure}

To further analyze the density dependence of the symmetry energy at
the saturation density, the values of the symmetry energy at
$\rho_c$ and $\rho_0$, the slope parameters $L$, and the
neutron-skin thickness $\Delta R_{\rm np}$ of $^{208}$Pb are listed
in Table I. One sees that for the seven Skyrme forces, i.e. SkSC4,
SkSC1, v075, v080, v090, v105, MSk3, the corresponding slope
parameters $L$ are very small and even negative. In Ref.
\cite{Dut12}, it is thought that the BSk, SkSC and MSk families
under-predict both the symmetry energy and its derivative at the
saturation density. To further check these selected Skyrme forces
for the description of other physical quantities, we study the
neutron-skin thickness of $^{208}$Pb. The linear relationship
between the slope parameter $L$ and the $\Delta R_{\rm np}$ of
$^{208}$Pb was observed in Refs.\cite{Cent09,Roca}. Fig. 6 shows the
slope parameter $L$ of these Skyrme forces as a function of the
corresponding neutron-skin thickness of $^{208}$Pb. The
corresponding values of $\Delta R_{\rm np} (^{208} {\rm Pb})$ from
the seven Skyrme forces with  $L<10$ MeV are about 0.11 fm. The
recent experimental measurements \cite{Klim07,Carb10,Tam11,Kras12}
for the $\Delta R_{\rm np}$ of $^{208}$Pb show $0.131 \leqslant
\Delta R_{\rm np} (^{208} {\rm Pb}) \leqslant 0.218 $ fm. It implies
that the seven Skyrme forces with small $L$ values are not suitable
for the description of the symmetry energy at around saturation
densities. The other ten Skyrme forces with $28 \leqslant L
\leqslant 65$ MeV, i.e., Ska25s20, Ska35s20, SV-sym32, SLy2, SLy6,
SLy7, SLy230a, BSk17, Skz1 and SkT7, reasonably well describe both
the Fermi energy difference and the neutron-skin thickness of
$^{208}$Pb. The obtained central value of $L$ with this approach is
generally consistent with the result from the neutron star mass and
radius measurements \cite{Stein}. Out of the ten forces, four with
the smallest average deviation [$\langle \sigma \rangle \leqslant
1.12$ MeV, see Table I] have values of $L=56\pm 9$ MeV for the slope
parameter, and the Skyrme forces Ska25s20, Ska35s20, SV-sym32 were
also recommended in the recent study of Ref.\cite{Dut12}.

\begin{figure}
     \centering
        \includegraphics[width=0.7\textwidth]{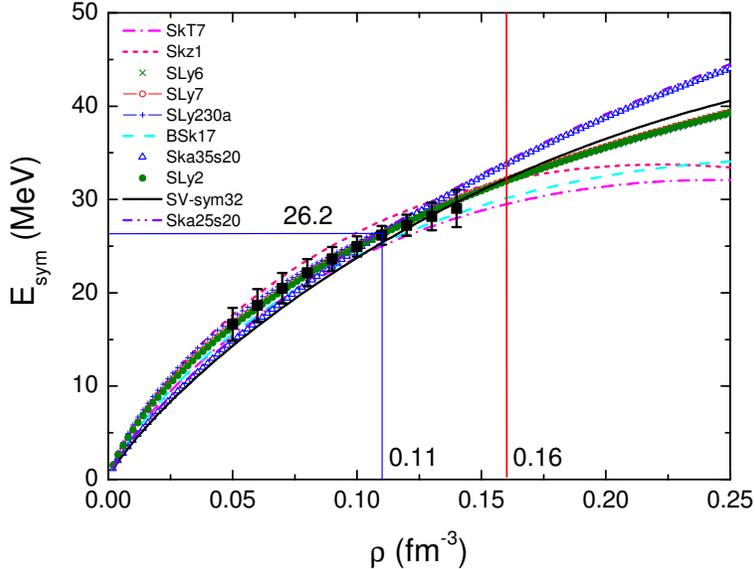}
        \caption{(Color online) Nuclear symmetry energy as a function of density. The squares denotes the results from the 17
selected Skyrme forces based on the Fermi energy difference. Other
symbols and curves denotes the results of ten selected Skyrme forces
with which both the Fermi energy difference and the neutron-skin
thickness of $^{208}$Pb can be reasonably well described.}
        \label{fig3}
    \end{figure}

\begin{table}
\caption{  Nuclear symmetry energy and neutron-skin thickness
$\Delta R_{\rm np}$ of $^{208}$Pb with the selected 17 Skyrme
forces. $\langle \sigma \rangle $ denote the calculated average
deviation according to Eq.(7) for the 19 nuclei. The unit of $\Delta
R_{\rm np}$ is fm, and those of others are MeV. The bold-face
entries denote the four forces with the smallest average deviation
and reasonable neutron-skin thickness of $^{208}$Pb. }
\begin{tabular}{ccccccc}
 \hline\hline
 Lable  & ~~$\langle \sigma \rangle $~~  & ~~$E_{\rm sym}(\rho_c)$~~    & ~~$E_{\rm sym}(\rho_0)$~~   & ~~$L$ ~~   & ~~$\Delta R_{\rm np} $ ($^{208}$Pb) ~~  & Reference   \\
\hline
\textbf{Ska25s20}     &  \textbf{1.05}   & $\textbf{26.6}$ &$\textbf{34.2}$ & $\textbf{65.1}$  & $\textbf{0.20}$ &  \cite{Dut12} \\
\textbf{SV-sym32}     &  \textbf{1.07}   & $\textbf{25.4}$ &$\textbf{32.1}$ & $\textbf{57.4}$  & $\textbf{0.19}$ &  \cite{SVsym32}\\
\textbf{ SLy2   }     &  \textbf{1.11}   & $\textbf{26.3}$ &$\textbf{32.1}$ & $\textbf{47.5}$  & $\textbf{0.16}$ &  \cite{SLy2}\\
\textbf{Ska35s20}     &  \textbf{1.12}   & $\textbf{26.3}$ &$\textbf{33.5}$ & $\textbf{64.4}$  & $\textbf{0.20}$ &  \cite{Dut12}\\
 BSk17        &  1.13   & $25.3$ &$30.0$ & $36.3$  & $0.15$ & \cite{HFB17} \\
 SLy230a      &  1.13   & $26.7$ &$32.0$ & $44.3$  & $0.15$ & \cite{SLy} \\
 SLy7         &  1.17   & $26.5$ &$32.0$ & $46.9$  & $0.16$ & \cite{Chab98}\\
 SLy6         &  1.18   & $26.4$ &$32.0$ & $47.5$  & $0.16$ & \cite{Chab98}\\
 Skz1         &  1.19   & $27.7$ &$32.0$ & $27.7$  & $0.15$ &  \cite{SKZ1}\\
 SkT7         &  1.19   & $25.0$ &$29.5$ & $31.1$  & $0.15$ &  \cite{SKT7}\\
              &         &        &       &         &        &  \\
SkSC4         &  0.97   & $27.1$ &$28.8$ & $-2.4$  & $0.11$ &  \cite{SKSC4}\\
v075          &  1.03   & $26.2$ &$28.0$ & $-0.3$  & $0.11$ &  \cite{V000}\\
v080          &  1.06   & $26.0$ &$28.0$ & $2.2$   & $0.11$ &  \cite{V000}\\
v090          &  1.14   & $25.8$ &$28.0$ & $5.1$   & $0.11$ &  \cite{V000}\\
MSk3          &  1.16   & $25.7$ &$28.0$ & $6.8$   & $0.11$ &  \cite{MSK3}\\
SkSC1         &  1.16   & $26.2$ &$28.1$ & $0.1$   & $0.11$ &  \cite{SKSC1}\\
v105          &  1.18   & $25.7$ &$28.0$ & $7.1$   & $0.11$ &  \cite{V000}\\
 \hline\hline
\end{tabular}
\end{table}

Fig. 7 shows the calculated nuclear symmetry energy as a function of
density. The squares (with $1.5\sigma$ uncertainty as the error bar)
denote the results from the 17 selected Skyrme forces listed in
Table I. Other symbols and curvatures denote the results from the
selected ten forces which can well describe the neutron-skin
thickness of $^{208}$Pb simultaneously. The symmetry energy at the
saturation density from the ten selected Skyrme forces is $E_{\rm
sym}(\rho_0)= 31.9\pm 2.1$ MeV with $1.5\sigma$ uncertainty.

\begin{center}
\textbf{IV. SUMMARY}
\end{center}

The correlation between the neutron-proton Fermi energy difference
$\Delta \varepsilon$ and the separation energy difference $\Delta S$
for some doubly-magic and semi-magic nuclei is analyzed with the
Skyrme energy density functionals and nuclear masses, with which
nuclear symmetry energy at sub-saturation densities is constrained
from 54 different Skyrme forces. The experimental data and the
Skyrme Hartreee-Fock calculations show $\Delta S \simeq \Delta
\varepsilon$ for closed-shell nuclei, even at the extremely
neutron-rich and proton-rich cases. The correlation between $\Delta
\varepsilon$ and $\Delta S$ is a good observation for studying the
nuclear symmetry energy at sub-saturation densities which probes the
isospin-dependent part of the Skyrme interaction, since the
cancelation between protons and neutrons directly removes the
influence of isospin-independent terms. The extracted symmetry
energy from 17 selected Skryme forces at the density of 0.11
fm$^{-3}$ is about 26.2 $\pm$ 1.0 MeV. The slope parameter of
symmetry energy is also studied by further combining the
neutron-skin thickness of $^{208}$Pb. Out of 54 Skyrme forces, ten
with $28 \leqslant L \leqslant 65$ MeV can reasonably well describe
both the Fermi energy difference and the neutron-skin thickness of
$^{208}$Pb. Within the ten forces, four Skyrme forces with the
smallest deviation (i.e. Ska25s20, Ska35s20, SV-sym32 and SLy2) have
values of $L=56\pm 9$ MeV for the slope parameter. The structures
and masses of finite nuclei at the ground states is helpful to
obtain the information of symmetry energy at the sub-saturation
densities. One should also note that the uncertainty of the symmetry
energy extracted from nuclear structures significantly increases
with the density at the region $\rho > 0.16$ fm$^{-3}$. It implies
that more isospin-sensitive observations from heavy-ion collisions
at intermediate and high energies and neutron stars are still
required for further constraining the symmetry energy at the
saturation and super-saturation densities.

\begin{acknowledgments}
We thank Professor Zhu-Xia Li for a careful reading of the
manuscript. This work was supported by the National Natural Science
Foundation of China Nos. 11275052, 11005022.
\end{acknowledgments}

\end{document}